\documentstyle{elsart}
\begin{document}
\begin{frontmatter}

\title{Stochastic Energetics of \\ Non-uniform Temperature Systems}
\author[m]{Miki Matsuo}
\author[s]{and Shin-ichi Sasa}
\address{Department of Pure and Applied Sciences, University of Tokyo. \\
Komaba, Meguro-ku, Tokyo 153}
\thanks[m]{e-mail: miki@jiro.c.u-tokyo.ac.jp}
\thanks[s]{e-mail: sasa@jiro.c.u-tokyo.ac.jp}
\begin{abstract}
We propose an energetic interpretation of
stochastic processes
described by Langevin equations with non-uniform temperature.
In order to avoid It\^{o}-Stratonovich dilemma, we start with a Kramers equation, and derive a Fokker-Plank equation by
the renormalization group method.
We give a proper definition of heat for the system.
Based on our formulations,
we analyze two examples, 
the Thomson effect and a Brownian motor 
which realizes the Carnot efficiency.
\end{abstract}
\begin{keyword}
non-uniform temperature system; Kramers equation; Fokker-Plank equation; stochastic energetics
\end{keyword}

\end{frontmatter}

\section{Introduction}

Recently, Brownian motors have attracted considerable attention 
stimulated by research on molecular motors.
A Brownian motor, which appeared in Feynman's famous textbook 
for the first time as a thermal ratchet~\cite{Feynman}, is the machine which can rectify thermal fluctuations 
and product a directed current. 
Until now, several models of a Brownian motor have been proposed. 
The most received and extensively researched system consists 
of Brownian particles subjected to a Gaussian white noise 
whose intensity is characterized by a temperature for the environment~\cite{Magnasco}.  
There are other types of Brownian motor models. In particular, 
it is known that a Brownian particle under non-uniform temperature 
profile moves directedly on the average 
when the potential profile synchronizes the temperature profile but they are not same shapes~\cite{Buttiker}.
Here, the motion of the particle was assumed to be described 
by Langevin dynamics supplemented with a state dependent noise. 

Until now, the energetic consideration for a Langevin dynamics with uniform temperature 
has been developed~\cite{Seki,Hondo1}. 
In fact, the efficiency for the Feynman's ratchet was
discussed based on the notions of heat and work which are
defined properly in a consistent way with thermodynamic laws.
Then, the question we address here is how to define heat and work
in non-uniform temperature systems. 
Thus, our aim of this paper is to extend 
the energetic consideration so as to apply to such systems. 
It will turn out that this question is not trivially
solved owing to the nature of the state dependent noise. 

In this paper, so as to avoid the complications for the state dependent noise,
we first derive a Fokker-Plank equation from a uniquely defined Kramers equation 
by employing the perturbative expansion.
There are many perturbation methods
to derive a Fokker-Plank equation~\cite{Bocquet,Kampen1}. Since we treat an delicate problem, the most formal
theory without heuristicity might be best. Thus, we adopt renormalization
group (RG) method, whose applicability to differential equations has been established recently~\cite{Oono3}.
Then, following the Sekimoto's formulation~\cite{Seki},
we make an energetic interpretation of the system. 
After that, as applications,
two kinds of non-uniform temperature systems are analyzed.
As the first example,
we study the Thomson effect for non-interacting particles
and show that the Thomson coefficient is $1/2$.
As the second, we calculate the work efficiency for the B\"{u}ttiker's model
and show that the Carnot limit is realized under the quasistatic condition.

This article is organized as follows.
In section2, we review a sort of dilemma 
caused by the nature of a state dependent noise.
In section3, the Fokker-Plank equation is derived from the Kramers equation 
by employing the renormalization group method.
In section4, we define heat for the system described by the Fokker-Plank equation.
In section5,
we analyze above two examples.
Section6 is devoted to a summary and conclusions of this work.


\section{It\^{o}-Stratonovich dilemma}

In this section, we review a dilemma caused by ill-defined
nature of a Langevin equation with a position dependent noise intensity~\cite{Risken}

\begin{equation}
\gamma \dot x = f + \sqrt{T(x)} \xi,
\label{overdamped}
\end{equation}

\begin{equation}
\langle \xi(t) \xi(t') \rangle = 2 \gamma \delta(t - t')\ ,
\end{equation}

where $\gamma$ is a friction constant of a Brownian particle,
$f$ is an external force, $T(x)$ is a temperature profile and 
we have assumed that
a Boltzmann constant $k$ is equal to the unity.
Owing to space dependence of a noise intensity,
the following difficulty
arises.
Because the noise has no correlation time, 
it is not yet clear which $x$ value one has to use in the 
function $T(x)^{1/2}$ in Eq.(\ref{overdamped}).
If, for instance, $\xi$ is considered as a sum of peaked functions
with no width, the stochastic variable $x$ will jump at every time
when such a peaked function occurs.
The question then arises: which
$x$ value must use in $T(x)^{1/2}$?
One may use the value $x$ just before the jump, after the jump, or some value between these two values.
From a purely mathematical point one cannot answer this question,
but one has to use a prescription as additional specification.
Among some prescriptions,
there are two representatives
called It\^{o} and Stratonovich.  
Under the It\^{o} prescription,
the value of $x$ is inserted before the jump, while
the mean value between before and after the jump is opted for, under the Stratonovich prescription.
The Fokker-Plank formalism explicitly shows this prescription dependence as follows:
A Fokker-Plank equation with the Stratonovich prescription is written as

\begin{equation}
\frac{\partial \phi}{\partial t} = \frac{1}{\gamma} \frac{\partial}{\partial x} \Bigl( T \frac{\partial}{\partial x} - f + \frac{1}{2} \frac{\partial T}{\partial x} \Bigl) \phi ,
\label{ito-fokker-plank}
\end{equation}

while a Fokker-Plank equation with the It\^{o} prescription is written as

\begin{equation}
\frac{\partial \phi}{\partial t} = \frac{1}{\gamma} \frac{\partial}{\partial x} \Bigl( T \frac{\partial}{\partial x} - f +  \frac{\partial T}{\partial x} \Bigl) \phi .
\label{stratonovich-fokker-plank}
\end{equation}

Such a dependence of an apparently unique equation on prescriptions 
is called "It\^{o}-Stratonovich dilemma", and
their difference 

\begin{equation}
\frac{1}{\gamma} \frac{\partial}{\partial x} \Bigl( \frac{1}{2} \frac{\partial T}{\partial x} \Bigl) \phi
\end{equation}

is called a spurious flow.
Owing to the existence of this dilemma,
we cannot proceed to discuss energetics,
because descriptions of energetics are subject to prescriptions.

However, it is possible to avoid this dilemma by considering the effect of inertia of a Brownian particle. An underdamped type of a Langevin equation is written as

\begin{equation}
\left\{
\begin{array}{l}
\dot x = v ,\\
\dot v = - \gamma v + f + \sqrt{T(x)} \xi ,
\end{array}
\right.
\label{underdamped}
\end{equation}

\begin{equation}
\langle \xi(t) \xi(t') \rangle = 2 \gamma  \delta(t - t'),
\end{equation}

where we have assumed the mass of a Brownian particle 
$m$ is equal to the unity.
From Eq.(\ref{underdamped}), we see that  
each delta peak in $\xi(t)$ causes the jump in $v(t)$, but no jump in $x(t)$ and, therefore, $T(x)^{1/2}$.
Since $\xi(t)$ does not cause a jump in the noise intensity,
the solution is uniquely determined for the given $\xi(t)$.
As a result, 
the evolution of a distribution function $P$ is uniquely 
denoted by

\begin{equation}
\Bigl( \frac{\partial}{\partial t} + v \frac{\partial}{\partial x}  + f \frac{\partial}{\partial v} \Bigl) P = \gamma \, \frac{\partial}{\partial v} \Bigl( v + T(x) \frac{\partial}{\partial v} \Bigl) P,
\label{kramers}
\end{equation}

which is called a Kramers equation.
Since there is no complication for the prescription dependence,
we discuss Eqs.(\ref{underdamped}) and (\ref{kramers}).

 
\section{Perturbation and Renormalization}\label{PandR}

In this section, employing a renormalization group method,
we derive a Fokker-Plank equation from  a Kramers equation 
with non-uniform temperature. 
Now we restrict ourselves to the one dimensional case so as to 
simplify the analysis. 
First, we rewrite Eq.(\ref{kramers}) as 

\begin{equation}
\frac{\partial}{\partial v} \Bigl( v + T(x) \frac{\partial}{\partial v} \Bigl) P  = \frac{1}{\gamma} \Bigl( \frac{\partial}{\partial t} + v \frac{\partial}{\partial x}  + f \frac{\partial}{\partial v} \Bigl) P.
\end{equation}

We consider a case that $\gamma$ is sufficiently large,
that is, a constant $1 / \gamma$ is regarded as a small parameter.
To emphasize smallness of the constant,  we rewrite 
$1 / \gamma$ as $\varepsilon$. Also, for the later convenience,
we rescale the time variable $t$ to $\varepsilon t$. The resultant
equation takes the form 

\begin{equation}
\frac{\partial}{\partial v} \Bigl( v + T(x) \frac{\partial}{\partial v} \Bigl) P -   \frac{\partial}{\partial t}P  = \varepsilon \Bigl(  v \frac{\partial}{\partial x} + f \frac{\partial}{ \partial v} \Bigl) P.
\label{kra}
\end{equation}

This rewriting of the time variable is not essential to the problem,
but necessary for us to employ 
the renormalization group method properly.

Now we seek  solutions under the condition $\varepsilon \rightarrow 0$.
We first expand the distribution function with $\varepsilon$
in such a way that  
\begin{equation}
P(x,v,t) \sim P^{(0)}(x,v,t) + \varepsilon P^{(1)}(x,v,t) + \cdots.
\end{equation}
Identifying terms of the same order, we obtain
\begin{eqnarray}
O(\varepsilon^0) &:& \frac{\partial}{\partial v} \Bigl( v + T(x) \frac{\partial}{\partial v} \Bigl) P^{(0)} -   \frac{\partial}{\partial t}P^{(0)}  = 0 \label{order0}, \\
O(\varepsilon^1) &:& \frac{\partial}{\partial v} \Bigl( v + T(x) \frac{\partial}{\partial v} \Bigl) P^{(1)} -  \frac{\partial}{\partial t}P^{(1)}  =  \Bigl(  v \frac{\partial}{\partial x} + f \frac{\partial}{\partial v} \Bigl) P^{(0)} \label{order1}, \\
O(\varepsilon^2) &:& \frac{\partial}{\partial v} \Bigl( v + T(x) \frac{\partial}{\partial v} \Bigl) P^{(2)} -  \frac{\partial}{\partial t} P^{(2)}  =  \Bigl(
 v \frac{\partial}{\partial x} + f \frac{\partial}{\partial v} \Bigl) P^{(1)}
\label{order2}, \\
&& \hspace{0.5cm} \cdots \cdots \nonumber.
\end{eqnarray}

We may successively solve these equations from  lower order.
As preparation for solving them,  we  define the Fokker-Plank 
operator as
 
\begin{equation}
\hat L_{FP} \equiv \frac{\partial}{\partial v} \Bigl( v + T \frac{\partial}{\partial v} \Bigl),
\end{equation}

and introduce the eigenfunctions of the operator, which
are written in the  form

\begin{equation}
L_n(x,v) = \frac{1}{\sqrt{T(x)}} H_n( \frac{v}{ \sqrt{2 T(x)}} )
\exp ( - \frac{v^2}{2 T(x)} ) ,
\end{equation}

where $H_n$ is the n-th Hermite polynomial.
The characteristic equation is given by

\begin{equation}
\hat L_{FP} L_n(x,v) = - n L_n(x,v).
\end{equation}

We also expand $P^{(m)}$ in terms of  eigenfunctions:

\begin{equation}
P^{(m)}(x,v,t) = \sum_{n=0}^{\infty} a_n^{(m)}(x,t) L_n(x,v).
\end{equation}

Now, let us solve Eq.(\ref{order0}).
Decomposing left-handed-side of Eq.(\ref{order0}) in terms of the eigenfunctions, 
we obtain 
\begin{equation}
\frac{\partial}{\partial t} a_n^{(0)}(x,t) = -na_n^{(0)}. 
\end{equation}
We notice that $a_n$, $(n \not= 0)$, decays exponentially in time.
Thus, discarding such terms,  we obtain
\begin{equation}
P^{(0)} = \frac{\phi(x)}{\sqrt{T(x)}}  \exp ( - \frac{v^2}{2 T(x)} ),
\label{p0}
\end{equation}
where $\phi(x)$ is an arbitrary function in $x$.

The physical picture behind this equation is as follows : 
In the fast time scale $t \simeq 1$, the velocity distribution relaxes to the Maxwellian rapidly
and space distribution stays just as it is.

We next proceed to calculate the higher order terms.
Substituting the  solution Eq.(\ref{p0}) into
Eq.(\ref{order1}), we obtain 

\begin{eqnarray}
\hat L_{FP} P^{(1)} -   \frac{\partial}{\partial t}P^{(1)} &=& \frac{T' \phi}{2 T^2} \frac{v^3}{\sqrt T} e^{- \frac{v^2}{2 T}}  + \Bigl( \frac{\partial \phi}{\partial x} - \frac{f \phi}{T} - \frac{T' \phi}{2 T}\Bigl) 
\frac{v}{\sqrt{T}} e^{- \frac{v^2}{2 T}} \\
&=& \frac{\sqrt{2 T}}{8 T} T' \phi L_3 +  \frac{\sqrt{2 T}}{2} \Bigl( \frac{\partial \phi}{\partial x} - \frac{f \phi}{T} + \frac{T'}{T} \phi \Bigl) L_1. \label{set}
\end{eqnarray}

where $T'$ denotes $d T / d x$. 
Eq.(\ref{set}) leads to a set of equations
\begin{eqnarray}
- 3 a_3^{(1)} - \frac{\partial}{\partial t} a_3^{(1)} &=& \frac{\sqrt{2 T}}{8 T} T' \phi ,\\
- a_1^{(1)} - \frac{\partial}{\partial t} a_1^{(1)}  &=& \frac{\sqrt{2 T}}{2} \Bigl( \frac{\partial \phi}{\partial x} - \frac{f \phi}{T} + \frac{T'}{T} \phi \Bigl) ,\\
-n a_n^{(1)} - \frac{\partial}{\partial t} a_n^{(1)} &=& 0 \ \ \ (n \neq 1,3).
\end{eqnarray} 

Neglecting terms decaying exponentially in time,
we obtain 

\begin{eqnarray}
P^{(1)} &=& - \frac{1}{3} \frac{\sqrt{2 T}}{8 T} T' \phi L_3 - \frac{\sqrt{2 T}}{2} \Bigl( \frac{\partial \phi}{\partial x} - \frac{f \phi}{T} + \frac{T'}{T} \phi \Bigl) L_1 \\
&=& - \frac{T' \phi}{6 T^2} \frac{v^3}{\sqrt{T}} e^{- \frac{v^2}{2 T}} - \Bigl( \frac{\partial \phi}{\partial x} - \frac{f \phi}{T} + \frac{T'}{2 T} \phi \Bigl) \frac{v}{\sqrt{T}} e^{- \frac{v^2}{2 T}}.
\label{lllsls}
\end{eqnarray}

Then we substitute $P^{(1)}$ 
into the $O(\varepsilon^2)$ equation so that
the right-handed-side of the equation takes the form
 
\begin{eqnarray}
\Bigl(  v \frac{\partial}{\partial x} + f \frac{\partial}{\partial v} \Bigl) P^{(2)} &=& - f \Bigl( \frac{\partial \phi}{\partial x} - \frac{f \phi}{T} + \frac{T'}{2 T} \phi \Bigl) \frac{1}{\sqrt{T}} e^{ - \frac{v^2}{2 T} } \nonumber \\
&& - \frac{\partial}{\partial x} \Bigl( \frac{\partial \phi}{\partial x} - \frac{f \phi}{T} + \frac{T'}{2 T} \phi \Bigl) \frac{v^2}{\sqrt{T}} e^{ - \frac{v^2}{2 T} } \nonumber \\
&& + \frac{T'}{2 T} \Bigl( \frac{\partial \phi}{\partial x} - \frac{f \phi}{T} + \frac{T'}{2 T} \phi \Bigl) \frac{v^2}{\sqrt{T}} e^{ - \frac{v^2}{2 T} } \label{0co} \\
&& - \frac{T' \phi f}{ 2 T^2 } \frac{v^2}{\sqrt{T}} e^{ - \frac{v^2}{2 T} } + \frac{f}{T} \Bigl( \frac{\partial \phi}{\partial x} - \frac{f \phi}{T} + \frac{T'}{2 T} \phi \Bigl) \frac{v^2}{\sqrt{T}} e^{ - \frac{v^2}{2 T} } \nonumber \\
&& - \frac{\partial}{\partial x} \Bigl( \frac{T' \phi}{6 T^2} \Bigl) \frac{v^4}{\sqrt{T}} e^{ - \frac{v^2}{2 T} }- \frac{T'}{2 T^2} \Bigl( \frac{\partial \phi}{\partial x} - \frac{f \phi}{T} + \frac{T'}{2 T} \phi \Bigl) \frac{v^4}{\sqrt{T}}
 e^{ - \frac{v^2}{2 T} } \nonumber \\
&& + \frac{ T' \phi f }{6 T^3} \frac{v^4}{\sqrt{T}} e^{ - \frac{v^2}{2 T} } + \frac{T'^2 \phi}{12 T^3} \frac{v^4}{\sqrt{T}} e^{ - \frac{v^2}{2 T} } - \frac{T'^2 \phi}{12 T^4} \frac{v^6}{\sqrt{T}} e^{ - \frac{v^2}{2 T} } \nonumber . 
\end{eqnarray}

In the same way as before, we derive equations for $a_n^{(2)}$.
In particular, $a_0^{(2)}$ satisfies 

\begin{equation}
- \frac{\partial}{\partial t} a_0^{(2)} = \psi,
\end{equation}

where

\begin{eqnarray}
\psi &\equiv& \int \Bigl(  v \frac{\partial}{\partial x} + f \frac{\partial}{\partial v} \Bigl) P^{(2)}\,  d\,v  \\
&=& - \frac{\partial}{\partial x} \Bigl( T(x) \frac{\partial \phi}{\partial x} - f \phi + T' \phi \Bigl) .
\label{secularterm}
\end{eqnarray}

Therefore, $P^{(2)}$  is expressed as

\begin{equation}
P^{(2)} =   - \varepsilon^2 \psi t  \frac{1}{\sqrt{T}} 
e^{ - \frac{v^2}{2 T} } +  O(\varepsilon) \, ( regular \, term ),
\end{equation}
and the asymptotic expansion up to the second order
is written as
\begin{equation}
P = ( \phi(x)  - \varepsilon^2 \psi t ) \frac{1}{\sqrt{T}} e^{ - \frac{v^2}{2 T} } +  O(\varepsilon) \, ( regular \, term ).
\label{rewriting}
\end{equation}
This shows that $P$ has  a secular term which is
unbounded in time.  Owing to the existence of the secular term,
the perturbation is valid  only in a short time interval
satisfying $\varepsilon^2 t \ll 1$.
Improve the result of naive perturbation, 
we apply the renormalization group method. 
We introduce a time $\tau$ when the initial condition for $\phi$
is given. That is, we rewrite  Eq.(\ref{rewriting}) as

\begin{equation}
P = \bigl( \phi(x,\tau)  - \varepsilon^2 \psi (t - \tau) \bigl) \frac{1}{\sqrt{T}} e^{ - \frac{v^2}{2 T} } + O(\varepsilon) \, ( regular \, term ),
\end{equation}

Since  $P$ should not depend on the choice of $\tau$,
we impose the condition

\begin{equation}
\Bigl( \frac{\partial P}{\partial \tau} \Bigl)_{\tau = t}  = 0,
\end{equation}

which is called a renormalization group equation~\cite{Oono3}.
This yields

\begin{equation}
 \frac{\partial \phi(x,t)}{\partial t} +  \varepsilon^2 \psi = 0 .
\end{equation}

Restoring the time scale to the original one and
using Eq.(\ref{secularterm}),  we finally obtain 
\begin{equation}
 \frac{\partial \phi(x,t)}{\partial t} - \frac{1}{\gamma} \frac{\partial}{\partial x} \Bigl( T(x) \frac{\partial \phi}{\partial x} - f \phi + T' \phi \Bigl) = 0 .
\label{reduc}
\end{equation}


\section{Energetics}

Equation (\ref{reduc}) takes the same form as a Fokker-Plank equation given by Eq.(\ref{ito-fokker-plank}).
Therefore, an overdamped Langevin equation corresponding to Eq.(\ref{reduc}) is simply given by Eq.(\ref{overdamped}) when the It\^{o} prescription is assumed.
However, we notice that this simpleness dose not ensure the uniqueness of prescription. We can consider that Eq.(\ref{reduc}) corresponds
the Langevin equation given by 

\begin{equation}
\gamma \dot x = \sqrt{T(x)} \xi + f - \frac{1}{2} \frac{\partial T}{\partial x},
\label{aim}
\end{equation}

when the Stratonovich prescription is assumed. 
Although we have two apparently different forms of Langevin equations,
we restrict ourselves to the Stratonovich prescription because
the Stratonovich prescription is more suitable for description of
energetics, as we will see below.

The revision term which we call the third term of right-handed-side of Eq.(\ref{aim}) is relevant when one faces energetics.
Following Sekimito~\cite{Seki,Cond}, we define heats for the systems described
by an overdamped and an underdamped Langevin equation with a position dependent noise intensity
as the following stochastic integrals

\begin{equation}
Q_{{\rm over}} \equiv - \int \{ \gamma \dot x  - \sqrt{T(x)} \xi \} dx(t),
\label{overheat}
\end{equation}

\begin{equation}
Q_{{\rm under}} \equiv - \int \{ \gamma  v - \sqrt{T(x)} \xi \} dx(t),
\label{underheat}
\end{equation}

where the integral is assumed in the Stratonovich sense.
Substituting Eq.(\ref{aim}) to Eq.(\ref{overheat}), we obtain the decomposition of the heat
of the overdamped case

\begin{equation} 
Q_{{\rm over}} = Q_U + Q_T,
\end{equation}

\begin{eqnarray}
Q_U &\equiv& - \int f \, dx(t), \\ 
Q_T &\equiv& \int \frac{1}{2} \frac{\partial T}{\partial x} dx(t),
\end{eqnarray}

where 
$Q_U$ is the part which an external force contributes to
and $Q_T$ is the part which position dependent temperature contributes to.
On the other hand, substituting 
Eq.(\ref{underdamped}) to Eq.(\ref{underheat}), 
we get the decomposition of the heat of the underdamped case 

\begin{equation} 
Q_{{\rm under}} = Q_U + Q_A,
\end{equation}

\begin{equation} 
Q_A \equiv \int \dot v dx(t).
\label{qa}
\end{equation}

Here, we show that $Q_T$ can be derived as the first order contribution
to $Q_A$ for the high friction limit.
Taking an ensemble average about both-handed-sides of Eq.(\ref{qa}), we get

\begin{eqnarray}
\langle Q_A \rangle &=& \int \langle \dot v dx \rangle \\
&=&  \frac{1}{2} \int dt \frac{\partial}{\partial t} \langle v^2 \rangle \label{111} \\
&=&  \frac{1}{2} \int dt \frac{\partial}{\partial t} \int dx \int dv v^2 P(x,v,t) .
\end{eqnarray}

Neglecting anything but the lowest contribution,
we obtain

\begin{eqnarray}
\langle Q_A \rangle &\sim& \frac{1}{2} \int dt \frac{\partial}{\partial t} \int dx \int dv v^2 P^{(0)}(x,v,t) \\
&=& \frac{1}{2} \int dt \frac{\partial}{\partial t} \int dx T(x) \phi (x,t) \label{222} \\
&=& - \frac{1}{2} \int dt \int dx T(x) \partial_x J(x,t) \\
&=& \frac{1}{2} \int dt \int dx \frac{\partial T}{\partial x} J(x,t) - \frac{1}{2} \int dt \Bigl[ T(x) J(x,t) \Bigl]^{x_R}_{x_L}\\
&=& \langle Q_T \rangle - \frac{1}{2} \int dt \Bigl[ T(x) J(x,t) \Bigl]^{x_R}_{x_L},
\label{vs}
\end{eqnarray}

where the system has been assumed to be an interval $[x_L,x_R]$.
The surface term may be interpreted as the energy flowing from the outside
of the system.
Then, as far as the energy exported form the heat bath is concerned,
we have only to pay attention to the bulk term. 
Thus, we conclude that $Q_T$ is the lowest order contribution to $Q_A$,
that is,

\begin{equation}
Q_A \sim Q_T.
\label{at}
\end{equation}

We shortly account for the meaning of the revision term.
We saw in section3 that 
the velocity distribution function rapidly relaxes to the position dependent Maxwellian in the high friction case.
The averaged energy of each position $H$ is written as

\begin{eqnarray}
H(x,t) &\equiv& \int ( \frac{1}{2} v^2 + U(x) ) P(x,v,t) dv \\
&\sim&    \int ( \frac{1}{2} v^2 + U(x) ) \phi(x,t) L_0(x,v) dv \ \ \ (\gamma \rightarrow \infty) \\
&=& ( \frac{1}{2} T(x) + U(x) ) \phi(x,t) .
\end{eqnarray}

This equation implies that the kinetic energy contributes to the effective potential 
via the equipartition law.
Thus, the effective potential may lead to the revision term.


\section{Examples}

In this section, we investigate two 
examples as application of the energetic interpretation made in section4.

\vspace{0.5cm}

{\bf Ex. 1) Thomson effect}

The Thomson effect, one of the thermoelectric effect, 
is the phenomenon that some heat is generated besides 
Joule heat when a current flows 
in a conductor at which a temperature gradient is charged.
The generated heat is called Thomson heat.
According to phenomenological laws, Thomson heat generated per unit time between two points of the conductor, written as $\dot Q$, takes the form

\begin{equation}
\dot Q = \tau J \Delta T, \label{pheno}
\end{equation}
 
where $\Delta T$ is the temperature difference between 
two points, $J$ is a current and $\tau$ is a material constant called
"Thomson coefficient"~\cite{Callen}.

Now let us calculate Thomson heat from a kinematic point of view.
We assume the system in question is an open interval $I=(0,1)$ on an real axis
and particle reservoirs are  points $ 0,1 $.
Next we give a few conditions.
We assume that the particle number distribution is realized with $P_0, P_1$ in each reservoirs, temperature $T(x)$ is externally given and there is no mechanical potential.
Using these conditions and Eq.(\ref{reduc}),
the problem we solve is written as

\begin{eqnarray}
\frac{\partial P}{\partial t} &=& \frac{\partial^2}{\partial x^2} \Bigl( T(x) P \Bigl) \label{glk} ,\\
P(0) &=& P_0 \, \  P(1) = P_1 .
\end{eqnarray}

Now we restrict ourselves to a steady state.
From Eq.(\ref{glk}), a steady state distribution $P_s$ satisfies the relation 

\begin{equation}
\frac{\partial}{\partial x} T P_s = - J \ ,
\label{suo}
\end{equation}

where $J$ means a particle number flux and is a 
constant whose value is decided by boundary conditions.
The solution of Eq.(\ref{suo}) takes the following form

\begin{eqnarray}
T(x) P_s(x) &=& - J x + T(0) P_0 \ \ ( 0 \le x \le 1 ), \\
J &=& T(0) P_0 - T(1) P_1 \ .
\end{eqnarray}

The heat generated in the system is given by

\begin{eqnarray}
Q &=& \frac{1}{2} \int \frac{\partial T}{\partial x} dx(t).
\label{ttt}
\end{eqnarray} 

We take an ensemble average for both-handed-sides of Eq.(\ref{ttt}), and then we obtain

\begin{eqnarray}
\langle Q \rangle &=& \frac{1}{2} \int dt \int^1_0 dx \frac{\partial T}{\partial x} J ,\\
&=& \frac{1}{2} \Delta T J t, 
\label{ppo}
\end{eqnarray}

where $\Delta T$ is defined by

\begin{equation}
\Delta T \equiv T(1) - T(0),
\end{equation}

and we have used the fact that $J$ is a constant.
Differentiating both-handed-sides of Eq.(\ref{ppo}) with $t$, we 
finally obtain

\begin{equation}
\langle \dot Q \rangle = \frac{1}{2} \Delta T J.
\end{equation}

Comparing this relation with the the phenomenological law Eq.(\ref{pheno}),
we obtain

\begin{equation}
\tau = \frac{1}{2} \ .
\end{equation}

Notice that, in our model, interactions between particles are not considered.
If the effect of the interactions are taken into account, the value of Thomson coefficient
would be shifted from 1/2.

\vspace{0.5cm}

{\bf Ex. 2) B\"{u}ttiker's model}

B\"{u}ttiker derived a rigorous steady solution of Fokker-Plank equations with a position dependent temperature~\cite{Buttiker}. 
Let us first review his results by employing Eq.(\ref{reduc}).
It is assumed
that the system is defined over an interval $[0,h]$ with periodic boundary conditions.
It is also assumed that
the force $f$ is given by

\begin{equation}
f = - \frac{\partial U}{\partial x}.
\end{equation}
 
This model has a steady current solution expressed by

\begin{equation}
J = Z \Bigl( 1 - \exp \{ \int^h_0 \frac{1}{T(x)} \frac{\partial U}{\partial x} dx \} \Bigl) ,
\label{current}
\end{equation}

where Z is a normalization constant~\cite{Buttiker}.
Equation (\ref{current}) implies that a finite current exists
when $U(x)$ and $T(x)$ satisfy the condition

\begin{equation} 
\int^h_0 \frac{1}{T(x)} \frac{\partial U}{\partial x} dx \neq 0.
\end{equation}

Now, in order to extract the work from this system,
a load is superimposed.
That is, the form of the mechanical potential is assumed as

\begin{equation}
U(x) = U_0(x) + L x \ \ (L > 0),
\end{equation} 

where $U_0(x)$ is a periodic function with a period $h$,

\begin{equation}
U_0(x+h) = U_0(x),
\end{equation}

and $L$ corresponds to the load per unit length.
When the temperaute $T$ and a periodic part of the potential $U_0$ satisfy
the inequality

\begin{equation} 
\int^h_0 \frac{1}{T(x)} \frac{\partial U_0}{\partial x} dx < 0,
\end{equation}

a Brownian particle climbs the slope of potential
so that it works against the load.
For simplicity, we assume the temperature function as

\begin{equation}
T(x)=
\left\{
\begin{array}{l}
T_1 \ \ : x \in [0,a] \\
T_2 \ \ : x \in [a,h] ,
\end{array}
\right.
\end{equation}

where $T_1 > T_2$.
Let us calculate the work efficiency 
defined by

\begin{equation}
\eta \equiv \frac{\langle W \rangle}{\langle Q _H \rangle},
\label{effdef}
\end{equation}

where $\langle Q_H \rangle$ is averaged heat extracted from the hotter heat bath whose temperature is $T_1$, and $\langle W \rangle$ is averaged work done by the particle against the load.
Since the hotter heat bath is located in the interval $[0,a]$,
$Q_H$ is written as 

\begin{equation}
Q_H = \int_{x(t) \in [0,a]} \Bigl\{ \frac{\partial U}{\partial x} + \frac{1}{2} \frac{\partial T}{\partial x} \Bigl\} dx(t).
\end{equation}

Taking an ensemble average, we get the denominator of Eq.(\ref{effdef})

\begin{eqnarray}
\langle Q_H \rangle &=& \int dt \int_0^a dx \frac{\partial U}{\partial x} J ,\\
&=& \{ U(a) - U(0) \} J t.
\label{QQQ}
\end{eqnarray}

The work against the load is defined by

\begin{equation}
W = L \int dx(t).
\end{equation}

(For more rigorous definition of work,
see \cite{Seki} and \cite{SekiSasa}.)
Then we can calculate the numerator Eq.(\ref{effdef}) as

\begin{eqnarray}
\langle W \rangle &=& L \langle \int dx(t) \rangle ,\\
&=& L h J t ,\\
&=& \{ U(h) - U(0) \} J t.
\label{WWW}
\end{eqnarray}

From Eq.(\ref{QQQ}) and Eq.(\ref{WWW}), the efficiency is expressed as

\begin{equation}
\eta =\frac{U(h) - U(0)}{U(a) - U(0)}.
\label{effi}
\end{equation}

We further notice that $U(h)$ and $U(a)$ are related to $T_1$ and $T_2$
for the quasistatic limit $J\rightarrow 0$.
In fact, using Eq.(\ref{current}), we get the relation 

\begin{equation}
\frac{U(a) - U(0)}{T_1} + \frac{U(h) - U(a)}{T_2} = 0.
\label{quasi}
\end{equation}

From Eq.(\ref{effi}) and Eq.(\ref{quasi}),
we obtain the efficiency for the quasistatic limit

\begin{equation}
\eta = \frac{T_1 - T_2}{T_1}.
\end{equation}

This is equal to the Carnot efficiency.
Recently, Sakaguchi has derived the Carnot efficiency by analyzing
a model with stochastic boundary conditions~\cite{Sakaguchi}.
His model turns out to be equivalent to that the limit $(h - a) \rightarrow 0$ is taken in our model.

\vspace{0.5cm}

Lastly, we address a few comments about above two examples.
In example1 the revision term intrinsically contributes to the averaged heat,
while the revision term has no contribution in example2.
This difference is caused by whether or not periodic boundary conditions are assumed.
As a matter of fact, we can generally say that, in one dimensional system, once a steady state is formed, 
the revision term does not affect to the averaged heat under periodic boundary conditions.
Since the steady condition leads to constant flux by the equation of continuity, 
the averaged value of heat is given by

\begin{equation}
\langle Q \rangle = \frac{1}{2} J t \int \frac{\partial T}{\partial x} dx \ .
\label{int}
\end{equation}

This is certainly equal to zero when periodic boundary conditions are assumed.
However, this matter is limited to one dimensional systems.
When the dimension is higher than one, 

\begin{equation}
{\rm div} J = 0
\end{equation}

is derived from the equation of continuity.
In this case, the circulation $\omega$ is defined by

\begin{equation}
\omega = {\rm rot} J,
\end{equation}

which may play an important role in energetics~\cite{Hondo2}.
We expect that the revision term contributes to the heat coupled with the circulation.
It is a future problem to study energetics in high-dimensional non-uniform temperature systems.


\section{Summary and Conclusions}

In this paper,
we have proposed an energetic interpretation about 
a stochastic process described by a Langevin equation 
with non-uniform temperature.
We analyzed two examples as applications of the energetic interpretation.
As the first example, the Thomson coefficient
for non-interacting particles was proved to be $1/2$.
The second example concerns the B\"{u}ttiker's model.
We calculated the work efficiency and 
showed that the Carnot efficiency is realized 
under the quasistatic condition.

\ack
The authors gratefully acknowledge T. Hondo, T. Shibata, F. Sasaki,
K. Kaneko and K. Sekimoto 
for useful discussions and helpful comments.
This work was partly supported by grants form the Ministry of Education, Science, Sports and Culture of Japan, No. 09740305.



\end{document}